\documentclass[12pt, draftclsnofoot, onecolumn]{IEEEtran}
\usepackage{url,amsmath,graphicx}
 \usepackage{stfloats}
 \usepackage{eqparbox}
  \usepackage{array}
  \usepackage{fixltx2e}
  \usepackage{stfloats}
   \usepackage{cite}
   \usepackage{url}
\usepackage{fancyhdr}
\usepackage{cite}
\usepackage{graphicx}
\usepackage{psfrag}
\usepackage{url}
\usepackage{stfloats}
\usepackage{amsmath}
\usepackage{array}
\usepackage{fancyhdr}

\usepackage{array, makecell}

\usepackage{amssymb}
\usepackage{color}

\begin{document}
	\title{\huge Performance Evaluation and Diversity Analysis of RIS-Assisted Communications Over Generalized Fading Channels in the Presence of Phase Noise}
\author{ Im\`ene~Trigui, \textit{Member, IEEE}, Wessam Ajib,  \textit{Senior Member, IEEE},\\  Wei-Ping Zhu, \textit{Senior Member, IEEE},  and Marco~Di~Renzo, \textit{Fellow, IEEE}

\thanks{Im\'ene Trigui and Wessam Ajib are with the D\'epartement d'informatique,
Universit\'e du Qu\'ebec à Montr\'eal, Montr\'eal, QC H2L 2C4, Canada, e-mail:
trigui.imen@courrier.uqam.ca and ajib.wessam@uqam.ca.}
\thanks{Wei-Ping Zhu is with the Department of Electrical
and Computer Engineering, Concordia University, Montreal, QC H3G 1M8,
Canada, e-mail: weiping@ece.concordia.ca.}
\thanks{M. Di Renzo is with Universit\'e Paris-Saclay, CNRS and CentraleSup\'elec, Laboratoire des Signaux et Syst\`emes,  Gif-sur-Yvette, France. (e-mail: marco.direnzo@centralesupelec.fr).}
}

\maketitle


\begin{abstract}
In this paper, we develop a comprehensive theoretical framework for analyzing the performance of  reconfigurable intelligent surfaces (RISs)-assisted communication systems over generalized fading channels and in the presence of phase noise. To this end, we propose the Fox's H model as a unified fading distribution for a large number of  widely used generalized fading channels. In particular, we  derive a unified analytical framework for computing the outage probability and for estimating the achievable diversity order of RIS-aided systems in the presence of phase shifts that either are optimally configured or are impaired by phase noise.  The
resulting expressions are general, as they hold for an arbitrary
number of reflecting elements, and various channel fading and phase noise distributions. As far as the diversity order is concerned, notably, we introduce an asymptotic analytical framework for determining the diversity order in the absence of phase noise, as well as sufficient conditions based on upper bounds and lower bounds for ensuring that RIS-assisted systems achieve the full diversity order in the presence of phase noise. More specifically, if the absolute difference between pairs of phase errors is less than $\pi/2$, RIS-assisted communications achieve the full diversity order over independent fading channels, even in the presence of phase noise. The theoretical frameworks and findings are validated with the aid of Monte Carlo simulations.
\end{abstract}



\section{Introduction}
Contemporary wireless networks modeling and analysis are vibrant
topics that keep taking new dimensions in complexity, as researchers explore the potential of innovative breakthrough technologies to support the  upcoming Internet of Things (IoT) and 6G era \cite{6G}.  Among these emerging technologies,  reconfigurable intelligent  surfaces (RISs) \cite{MDR_Eurasip}-\!\!\!\cite{Qw} have been introduced  with an  overarching vision of artificially controlling the wireless environment for increasing the quality of service
and spectrum efficiency. RIS technology is based on the massive integration of low-cost tunable passive elements on large surfaces,  which can be deployed on, e.g., the facades of buildings, and are  able  to, e.g., reflect and modulate the incident RF signals, which leads to a more controllable wireless environment \cite{ren4} and a more efficient implementation of multi-antenna transmitters \cite{MetaModulation}. Leveraging  these key properties, RIS-enabled networks challenge  device–side approaches, such as massive multiple-input-multiple-output (MIMO) systems, encoding, modulation, and relaying, which are currently deployed in wireless networks in order to fully adapt to the time-variant and unpredictable channel states \cite{renzo}, \cite{ren4}. Due to the potential opportunities offered by RIS-empowered wireless networks, a large body
of research contributions have recently appeared in the literature. The interested readers are
referred to the survey papers in \cite{MDR_Eurasip}-\!\!\!\cite{Qw}, where a comprehensive description
of the state-of-the-art, the scientific challenges, the distinctive differences with other technologies,
and the open research issues are discussed.

\subsection{Related Works}
Several research papers have appeared recently, mostly considering application scenarios where the line-of-sight link is either too weak or is not available, and, therefore,
an RIS is employed to enable the reliable transmission of data through the optimization of the phase shifts of
its individual reconfigurable elements and of the precoding and decoding vectors at the transmitter and receiver,
respectively, e.g., \cite{Rui_TWC}-\!\!\cite{s8}.
Specifically, several RIS-aided designs  have been recently proposed for  various advanced communication techniques, including millimeter-wave communications \cite{s3}, unmanned aerial
vehicle networks \cite{s8}, physical layer security \cite{s5}, simultaneous wireless information and power transfer \cite{s1}. However, there exist limited research efforts that have explored the
communication-theoretic performance limits of RIS-assisted communications \cite{CL1}-\!\!\!\cite{M4}, and, therefore, a limited number of results are available to date.  A major research issue for analyzing the fundamental performance limits of RIS-aided systems is the analysis of the exact  distribution of the RIS end-to-end equivalent channel. To circumvent this open issue,  some recent  attempts for studying RIS-aided systems include the use of approximate distributions and asymptotic analysis \cite{CL1}-\!\!\cite{M4}.   Under the assumption of Rayleigh fading, it was shown in \cite{CL1}, \cite{CL2}, \cite{CL3} that the
distribution of a single RIS equivalent channel follows a modified Bessel function. In \cite{CL4} and \cite{CL5}, an RIS-aided transmission system in the presence of phase errors was considered, and the
composite channel was shown to be equivalent to a point-to-point
Nakagami-m fading channel by using the central limit theorem (CLT).  It
is known, however, that the CLT is inaccurate when the number of reconfigurable
elements of the RIS is small. Recent results, in addition, showed that the
approximation error attributed to the CLT can be significant
in the high signal-to-noise-ratio (SNR) regime \cite{CL1}. Approximations for the received SNR in the presence of multiple randomly deployed RISs were introduced in \cite{CL6}, and  an asymptotic analysis of the data rate and channel
hardening effect in an RIS-aided large antenna-array system was
presented in \cite{CL7}. As far as the fading channel is concerned, the Rayleigh fading model has been commonly assumed, with only some exceptions that incorporated the line-of-sight (LoS) channel, yet only under the scope of phase optimization, e.g., \cite{nadeem}, \cite{zhang}. Rayleigh fading may  have, however, limited legitimacy in RIS-aided communications in which the RIS is appropriately deployed to leverage the LoS paths for enhancing the received power.

In Table I, we summarize the communication-theoretic frameworks that, to the best of our knowledge, are available in the literature. We evince that analytical studies have been conducted predominantly over Rayleigh fading channels and that no exact analytical framework exists. In addition, the analysis of the diversity order of RIS-assisted communication systems is still an open issue. There is, however, general consensus that the CLT is not a suitable tool for analyzing the diversity order, since it yields accurate approximations in the low-SNR regime \cite{CL1}. In this paper, we propose new analytical methods for overcoming these limitations.

\begin{table*}
\caption{PERFORMANCE ANALYSIS FRAMEWORKS AND DIVERSITY ANALYSIS AVAILABLE IN THE LITERATURE ($N$ = number of reconfigurable elements of the RIS)} 
\centering
\begin{tabular}{c|c|c|c|c|c}
  \hline
  \textbf{Ref.} & \textbf{Channel Model}  & \textbf{Phase Noise} & \textbf{Method of Analysis} & \makecell{\textbf{Diversity Order}\\ \textbf{(without phase noise)}} & \makecell{\textbf{Diversity Order}\\ \textbf{(with phase noise)}}\\ \hline
  \cite{bour} & Rayleigh  & No & Laguerre Series &    $\frac{N}{2}\frac{\pi^{2}}{16-\pi^{2}}$ (as for the CLT) &- \\
  \cite{CL2} & Rayleigh  & No & Gamma Approximation &$N$ & - \\
  \cite{CL3} & Rayleigh  & No & Method in \cite{gian} & $N$ & - \\
  \cite{CL4} & Arbitrary  & Yes & Large $N$, Nakagami Distr.  & $m$ in \cite[Eq. (12)]{CL4} & $m$ in \cite[Eq. (12)]{CL4} \\
  \cite{CL5} & Rayleigh  & $L$-bit Quant. & Bounds & $N$ & $<\frac{N+1}{2}$ if $L=2$; $N$ if $L \ge 3$ \\
  \cite{renzoquant1} & Rayleigh  & Yes & Large $N$, Gamma Approx. & $1/\rm{AF}$ in \cite[Eq. (20)]{renzoquant1} & $1/\rm{AF}$ in \cite[Eq. (20)]{renzoquant1} \\
  \cite{l1} & Rayleigh & No & ${\cal G K}$ Approximation & $N$ & - \\
  \cite{M1} & Rayleigh  & No & Chernoff Bound  & $N$ & - \\
  \cite{M2} &  Rayleigh  & $1$-bit Quant. & Gamma Approximation & $N$  & $({N+1})/{2}$ \\
  \cite{M3} & Rice  & No & Method in \cite{gian} & $N$  & -\\
  \cite{M4} & Rayleigh  & No & ${\cal G K}$ Approximation & $N$  & - \\
  This & Arbitrary Fox's H  & Arbitrary & Exact and Bounds & see (13)  & see Prop. 9 \\
\hline
\end{tabular} 
\end{table*}

\subsection{Contributions}
As a step forward to fill the mentioned research gaps, this work leverages fundamental results from Fox's H-transform theory for analyzing the performance of RIS-aided wireless communications. More precisely, we introduce a new analytical framework that provides exact analytical expressions of the outage probability  for several widely used generalized fading models in  the absence of phase noise.  The  proposed method for performance evaluation  is endowed with high flexibility to capture a broad range of fading distributions, thereby unveiling the diversity order of RIS-aided networks and generalizing the results available for transmission over Rayleigh fading. In addition, we introduce a new approach for analyzing the diversity order of RIS-aided systems over generalized fading channels in the presence of phase noise. The proposed approach confirms the unsuitability of the CLT for analyzing the diversity order of RIS-aided systems, and unveils the achievable diversity order under general fading channels and phase noise distributions. With the aid of lower bounds and upper bounds, more precisely, we identify sufficient conditions for achieving the full diversity order in RIS-assisted systems. The main
contributions of this paper can be summarized as follows.

\begin{itemize}
\item  We propose a new analytical framework for analyzing the performance of RIS-aided systems, which leverages Fox's H transform theory for modeling, in a unified fashion,  general RIS-induced fading environments in terms of outage probability and  achievable diversity.
\item  We draw multiple link-level design
insights from the proposed analysis. For instance, we show that the diversity order in the absence of phase noise scales with the number of reconfigurable elements of the RIS
multiplied by a factor that depends of the worst fading distribution of the
transmitter-RIS and RIS-receiver links.
\item  We study the transmission through an RIS whose phase shifts deviate from the ideal values according to general phase noise distributions, and discuss how the presence of errors in the phase
shifts influences the achievable diversity. We demonstrate, in particular, that the a sufficient condition for achieving the full diversity order is that the absolute difference  between  pairs of phase  errors is less than $\pi/2$.
\end{itemize}

The rest of this paper is organized as follows. Section II describes the system model and the
considered fading distributions. Sections III and IV are devoted to the  unified performance analysis framework,  where the  outage probability and the diversity order of RIS-assisted communications  are analyzed in the presence of perfect and imperfect phase shifts, respectively. Simulation and numerical results
are discussed in Section V. Finally, Section VI concludes the paper.

\section{System Model}
	We consider  an RIS with $N$ reconfigurable elements, which are arranged in a uniform array of tiny antennas spaced half of the wavelength apart and whose phase response is
locally optimized.  We assume that the RIS transmits data to a single antenna receiver by reflecting an incident RF wave emitted by a single antenna transmitter. More specifically, we assume that the direct transmission link between the transmitter and the receiver is blocked, and, thus, the RIS is deployed to relay the scattered signal and to leverage virtual LoS paths for enhancing the strength
of the received signal.  The received SNR of the considered system is  \cite{basar1}
\begin{equation}
\gamma=\rho\left|\sum_{i=1}^{N} h_ig_i e^{j \phi_i}\right|^{2},
\label{snr}
\end{equation}
where $\rho$ is the average SNR of the RIS-assisted link, $h_i$ and $g_i$, $i = 1, 2,\ldots,N$  are independent\footnote{The assumption of independent channel coefficients is made for analytical tractability and is justified, as a first-order approximation, if the reconfigurable elements of the RIS are spaced half of the wavelength apart. The generalization of the proposed analytical framework in the presence of channel correlation is postponed to future research works.} complex coefficients that characterize the channels between  the transmitter and the RIS, and the RIS and the receiver, respectively,  and $\phi_1, \ldots, \phi_N$  are the phase shifts that are optimized to  maximize the SNR at the receiver. In particular, $\rho$ in  (\ref{snr}) includes the path-loss of the end-to-end RIS channel, as described in, e.g., \cite{channel_1}-\!\!\cite{channel_3}, which is assumed to be fixed and given in this paper. The impact of channel estimation errors and overhead is not explicitly discussed in the present paper, but it can be taken into account as recently described in \cite{Overhead_1}.

\vspace{0.25cm}
\textit{Assumption 1:}  The amplitudes $\left|h_i\right|$ and
$\left|g_i\right|$ are independent and non-identically distributed (i.ni.d)   Fox's H-distributed random variables (RVs) whose  probability density function (pdf) is
\begin{equation}
f_{\mid y_i \mid}(x)= \kappa^{y}_{ i} {\rm H}_{p^{y}_{i},q^{y}_{i}}^{m^{y}_{i},n^{y}_{i}}\left[ c^{y}_{i} x \left|
\begin{array}{ccc} (a_{ij}, A_{ij})^{y}_{j=1: p^{y}_{i}} \\ (b_{ij},B_{ij})^{y}_{j=1: q^{y}_{i}} \end{array}\right. \right],
\label{h}
\end{equation}
where $y\in\left\{h ,g\right\}$, and ${\rm H}[\cdot]$ stands for the Fox's H function \cite[Eq. (1.2)]{mathai}.
 The Fox H distribution subsumes a  large number of  conventional and generalized fading distributions widely
 used in wireless communications, such as Rayleigh, Nakagami-m, and Weibull fading.

Usually, the RISs are positioned
to exploit the LoS path with respect to the location of the transmitter to increase the
received power. In this case, Rician fading is a better small-scale  fading model in the presence of a LoS path. However, the Rician distribution does not belong to the family of distributions in \eqref{h}. Therefore, we consider a further generalized fading model.

\vspace{0.25cm}
\textit{Assumption 2:} Using the  hyper-Fox H-distribution,  the pdf of a Rician fading channel is
\begin{equation}
f_{\mid y_i \mid}(x)=\sum_{k=0}^{\infty} \kappa^{y}_{ k} H_{0,1}^{1,0}\left[ c^{y}_{k} z \left|
\begin{array}{ccc} - \\ (k+\frac{1}{2},\frac{1}{2}) \end{array}\right. \right],
\label{rice}
  \end{equation}
where $c^{y}_{k}=\sqrt{K^{y}+1}$, $\kappa^{y}_{ k}=\frac{e^{-K^{y}}{K^{y}}^{k}\sqrt{K^{y}+1}}{\Gamma(k+1)}$, and $K^{y}$ for $y\in\left\{h ,g\right\}$ denotes the Rice factors of the transmitter-RIS and RIS-receiver links, respectively.

In the following sections, by leveraging the H-transform, we establish a unified framework for analyzing the performance  of RIS-assisted communications where the fading envelope is described by the Fox's H-distribution for non-specular small-scale fading and  the hyper Fox's H-distribution under LoS propagation. Perfect and imperfect phase shifts are analyzed.

\section{Outage Probability - No Phase Noise}

The optimal design for the phase shifts of an RIS-assisted link consists of setting the phase
shift of each element $\phi_i$ so that all phase contributions due to the phase of $h_i$, i.e., $\angle{h_i}$ and the phase of $g_i$, $ \angle{g_i}, i=1,\ldots, N$, are compensated \cite{basar1}. Accordingly,  substituting  $\phi_n=-\angle(h_n+g_n), n=1,\ldots, N$, in (\ref{snr}), the outage probability in the absence of phase noise is ${\rm P}\left(\rho\left(\sum_{i=1}^{N} |h_i||g_i|\right)^{\!2}<\gamma_{th}\right)={\rm P}\left(\sum_{i=1}^{N} |h_i||g_i|< \sqrt{\rho_t}\right)$, where $\rho_t={\gamma_{th}}/{\rho}$. An analytical expression of the outage probability is given in the following proposition.

\vspace{0.25cm}
\textit{Proposition 1:}
 The outage probability with optimal phase shifts is
\begin{eqnarray}
\!\! { \Pi}(\rho, N)=
\tau{\rm H}_{0,1:{\widetilde p}_1, {\widetilde q}_1,\ldots, {\widetilde p}_N,  {\widetilde q}_N}^{0,0:{\widetilde m}_1, {\widetilde n}_1,\ldots, {\widetilde m}_N,  {\widetilde n}_N} \hspace{2.75cm} \!\!\!\!\!\!\!\!\!\!\!\!\!\!\!\!\!\!\!\!\!\!\!\!\!\!\!\!\!\!\!\!\!\!\!\!\!\!\!\!\!\left[\!\!\!\begin{array}{ccc} {\widetilde{c}}_{1} \sqrt{\rho_t}\\ \vdots \\ {\widetilde{c}}_{N} \sqrt{\rho_t}\end{array} \!\!\left |\!\!\begin{array}{ccc}  -: (1,1), (\delta_{1},\Delta_{1})_{\widetilde{p}_{1}}; \ldots; (1,1), (\delta_{N},\Delta_{N})_{\widetilde{p}_{N}} \\(0;1,\ldots, 1): (\xi_{1},\Xi_{1})_{\widetilde{q}_{1}}; \ldots;  (\xi_{N},\Xi_{N})_{\widetilde{q}_{N}} \end{array}\right. \!\!\!\!\right],
\label{pout}
\end{eqnarray}
where $\tau=\prod_{i=1}^{N}\frac{\kappa^{h}_{i}\kappa^{g}_{i}}{{c^{h}_{i}c^{g}_{i}}}$, ${\widetilde{c}}_{i}=c^{h}_{i}c^{g}_{i}$, and $H[\cdot, \ldots, \cdot]$ is the multivariable Fox's H-function whose definition  in terms of Mellin-Barnes contour integrals is given in \cite[Definition A.1]{mathai} with
\begin{equation}
(\delta_{i},\Delta_{i})_{\widetilde{p}_{i}}=\left((a_{ij}+A_{ij},A_{ij})^{h}_{j=1:p^{h}_{i}}, (a_{ij}+A_{ij},A_{ij})^{g}_{j=1:p^{g}_{i}}\right)
 \end{equation}
\begin{equation}
(\xi_{i},\Xi_{i})_{\widetilde{q}_{i}}=\left((b_{ij}+B_{ij},B_{ij})^{h}_{j=1:q^{h}_{i}}, (b_{ij}+B_{ij},B_{ij})^{g}_{j=1:q^{g}_{i}}\right)
\end{equation}
and ${\widetilde m}_i=m^{h}_{i}+ m^{g}_{i}$, $ {\widetilde n}_i=n^{h}_{i}+ n^{g}_{i}+1$,  ${\widetilde q}_i=q^{h}_{i}+ q^{g}_{i}$, ${\widetilde p}_i=p^{h}_{i}+ p^{g}_{i}+1$.

\textit{Proof:} Let $\Gamma(\cdot)$ be the gamma function and $j$ be the imaginary unit.
By defining  ${\cal S}=\sum_{i=1}^{N}\left|h_i\right| \left|g_i\right|$, we have \cite{trigs}
\begin{equation}
{ \Pi}(\rho, N)=\frac{1}{2 \pi j}\int_{\cal C} s^{-1}\Psi_{\cal S}(s)e^{s \sqrt{ \rho}}ds,
\label{eq1}
\end{equation}
where $\Psi_{\cal S}(s)=\prod_{i=1}^{N}{\cal L}(f_{\left|h_i\right|\left|g_i\right|})(s)$ and ${\cal L}(\cdot)$ stands for the Laplace transform. Applying  \cite[Theorem (4.1)]{FOX} for the pdf of the product of two Fox's H distributions yields $f_{\left|h_i\right|\left|g_i\right|}$, which is, in turn, the Fox's H distribution as follows
\begin{eqnarray}
f_{ \left|h_i\right|\left|g_i\right|}(x)= \kappa^{h}_{ i}\kappa^{g}_{ i} \hspace{3cm} \!\!\!\!\!\!\!\!\!\!\!\!\!\!\!\!\!\!\!\!\!\!\!\!\!\!\!\!\!\!\!\!\!\! \!\!\!\!\!\!\!\!{\rm H}_{\widetilde{p}_{i}-1,\widetilde{q}_{i}}^{\widetilde{m}_{i},\widetilde{n}_{i}-1}\!\!\left[\! c^{h}_{i} c^{g}_{i} x\! \left|\!\!\!
\begin{array}{ccc} \left\{(a_{ij}, A_{ij})^{h}_{j=1: p^{h}_{i}},(a_{ij}, A_{ij})^{g}_{j=1: p^{g}_{i}} \right\} \\ \left\{(b_{ij},B_{ij})^{h}_{j=1: q^{h}_{i}},(b_{ij},B_{ij})^{g}_{j=1: q^{g}_{i}} \right\} \end{array}\right.\!\!\!\!\! \right].
\label{h}
\end{eqnarray}

Then, by evaluating the  Laplace transform of $f_{\left|h_i\right|\left|g_i\right|}$ with the help of \cite[Eq. (2.20)]{mathai}, $\Psi_{\cal S}$ follows as shown in (\ref{mgf}) at the top of this page.
\begin{figure*}[!t]
\begin{eqnarray}
{\Psi}_{\cal S}(s)&=&\frac{\tau}{(2\pi w)^{N}}\int_{{\cal C}_1} \ldots \int_{{\cal C}_N}\prod_{i=1}^{N}\left(\frac{\Gamma(-u_i)\Theta_i(u_i)}{{\widetilde{c}}_{i}^{u_i}}\right) s^{\sum_{i=1}^{N} u_i} du_1 du_2\ldots d_{u_N}\nonumber \\ &&\!\!\!\!\!\!\!\!\!\!\!\!\!\!
\text{where}\quad w=\sqrt{-1}, \quad
\Theta_j(u_j)=\frac{\prod_{j=1}^{{\widetilde m}_j}\Gamma\left(\xi_{j}+\Xi_{j}u_j)\right)\prod_{j=1}^{{\widetilde n}_j}\Gamma\left(1-\delta_{j}-\Delta_{j}u_j)\right)}{\prod_{j={\widetilde n}_j+1}^{{\widetilde p}_j}\Gamma\left(\delta_j+\Delta_{j}u_i\right)\prod_{j={\widetilde m}_j+1}^{{\widetilde q}_j}\Gamma(1-\xi_{j}-\Xi_{j}u_j)}\nonumber \\ &&
\hspace{-0.90cm} \text{and} ~~ {\cal C}_i, i=1,\ldots, N~~ \text{represents the contours} ~~ \left[\tau_i-w\infty,\tau_i+w\infty\right], \tau \in \mathbb{R}.\label{mgf}
\end{eqnarray}
\hrulefill 
\end{figure*}
By plugging (\ref{mgf}) into (\ref{eq1}), the outage probability can be written as in (10) shown at the top of the next page. The desired result follows by recalling  that $\frac{1}{2 \pi j}\int_{\cal L} s^{-a} e^{s  z}ds=\frac{z^{a-1}}{\Gamma(a)}$ and by capitalizing on the multiple Mellin–Barnes type contour integral of the multivariate Fox's H function \cite[Def. A.1]{mathai} with the aid of some algebraic manipulations.

\vspace{0.25cm}
\textit{Remark 1:} Although the numerical evaluation of the multivariate Fox's H-function is unavailable in
mathematical software packages, e.g., MATLAB and Mathematica, efficient implementations have been
reported in the literature \cite{Fimp}. They are used to obtain the numerical results.
\begin{figure*}
	\begin{eqnarray}
\frac{1}{2 \pi w}\int_{\cal L} s^{-1}\Psi_{\cal S}(s)e^{s  z}ds&=&\frac{\tau}{(2\pi w)^{N}}\!\int_{{\cal C}_1}\ldots \int_{{\cal C}_N}\prod_{i=1}^{N}\left(\frac{\Gamma(-u_i)\Theta_i(u_i)}{{\widetilde{c}}_{i}^{u_i}}\right)  \\ && \times \frac{1}{2 \pi w}\int_{\gamma+w\infty}^{\gamma-w\infty}e^{s z} s^{\sum_{i=1}^{N} u_i-1} ds du_1 du_2\ldots d_{u_N}\nonumber \\  && \hspace{-1.5cm} =  \frac{\tau}{(2\pi w)^{N}}\!\int_{{\cal C}_1}\ldots \int_{{\cal C}_N}\prod_{i=1}^{N}\left(\frac{\Gamma(-u_i)\Theta_i(u_i)}{{\widetilde{c}}_{i}^{u_i}}\right)\frac{z^{-\sum_{i=1}^{N} u_i}}{\Gamma(1-\sum_{i=1}^{N}u_i)}du_1 du_2\ldots d_{u_N}. \nonumber\label{LMGF}
	\end{eqnarray}
\hrulefill 
\end{figure*}
\vspace{0.25cm}
\textit{Proposition 2:} The outage probability of RIS-assisted communications in Rice fading is
\begin{eqnarray}
 { \Pi}(\rho, N)&=&\sum_{k_1, \ldots, k_N=0}^{\infty}\sum_{t_1, \ldots, t_N=0}^{\infty}
\tau{\rm H}_{0,1:1, 2,\ldots, 1, 2}^{0,0:2, 1,\ldots,2, 1}\Bigg[\!\!\!\begin{array}{ccc} \Delta \sqrt{\rho_t}\\ \vdots \\ \Delta \sqrt{\rho_t}\end{array} \!\!\Bigg| \nonumber \\ && \hspace{4cm} \!\!\!\!\!\!\!\!\!\!\!\!\!\!\!\!\!\!\!\!\!\!\!\!\!\!\!\!\!\!\!\!\!\!\!\!\!\!\!\!\!\!\!\!\!\!\!\!\!\!\begin{array}{ccc}  -: (1,1); \ldots; (1,1) \\(0;1,\ldots, 1)\!:\! (k_1+1,\frac{1}{2}), (t_1+1,\frac{1}{2}); \ldots;  (k_N+1, \frac{1}{2}),(t_N+1, \frac{1}{2})  \end{array}\Bigg. \Bigg],
\label{poutr}
\end{eqnarray}
where $\Delta=\sqrt{\left(1+K^{h}\right)\left(1+K^{g}\right)}$ and $\tau=\prod_{i=1}^{N}\frac{e^{-K^{h}}{K^{h}}^{k_i}}{\Gamma(k_i+1)}\frac{e^{-K^{g}}{K^{g}}^{t_i}}{\Gamma(t_i+1)}$.

\textit{Proof:} We use the distribution in (\ref{rice}) and apply  the same procedure as for the proof of (\ref{pout}).

\vspace{0.25cm}
\textit{Remark 2:}  The derived analytical expressions for the outage probability   in (\ref{pout}) and (\ref{poutr}) are general and new, and can be easily mapped into most  existing
 fading models. The obtained analytical frameworks are, to the best of our knowledge,  the first ones in the literature that yield the exact end-to-end SNR distribution of an RIS-assisted systems
 in terms of the multivariate Fox’s H-function. This is in contrast with
the recently reported expressions in \cite[Eqs. (4), (7)]{basar1} and \cite[Eq. (17)]{CL2}, which are based on approximations (the CLT in \cite{basar1} and the moment-based Gamma approximation in \cite{CL2}), in order to overcome the intricacy of the exact statistical modeling of the end-to-end SNR in RIS-aided systems. The novelty of the proposed approach is also apparent from the summary given in Table I. For the convenience of the readers, Table II provides the explicit expression of the outage probability for several widely used fading distributions.

\begin{table*}
\caption{OUTAGE  PROBABILITY OF RIS-ASSISTED SYSTEMS OVER
WIDELY USED FADING CHANNEL MODELS} 
\centering
\begin{tabular}{p{2.3in} p{4in}}
  \hline\hline
  \textbf{Instantaneous Fading Distribution} & \textbf{\quad \quad Outage Probability} $ {\Pi}(\rho, N)$ \\ \hline\hline
Nakagami-$m$ Fading \cite[Table IIV]{FoxH}:
$\begin{aligned} &f_{\mid y_i \mid}(x)=\frac{\sqrt{m^{y}_i}}{\Gamma(m^{y}_i)} {\rm H}_{0,1}^{1,0}\left[ \sqrt{m^{y}_i} x \left|
\begin{array}{ccc} - \\ (m^{y}_i-\frac{1}{2},\frac{1}{2}) \end{array}\right. \right]\end{aligned}$  & $\begin{aligned} \quad \quad &{\Pi}(\rho, N)=\left(\prod_{i=1}^{{N}} \Gamma(m^{h}_i)\Gamma(m^{g}_i)\right)^{-1}\\ &{\rm H}_{0,1:1,2,\ldots, 1, 2}^{0,0:2, 1,\ldots, 2, 1}\left[\!\!\!\begin{array}{ccc} \sqrt{m^{h}_1 m^{g}_1 }\sqrt{\rho_t}\\ \vdots \\ \sqrt{m^{h}_N m^{g}_N } \sqrt{\rho_t}\end{array} \!\!\left |\!\!\begin{array}{ccc}  -: (1,1), -; \ldots; (1,1), - \\(0;1,\ldots, 1): \{\xi_{1},\Xi_{1}\} ; \ldots;  \{\xi_{N},\Xi_{N}\} \end{array}\right. \!\!\!\!\right] \\ & \quad\quad \quad  \{\xi_{i},\Xi_{i}\}=(m^{h}_i, \frac{1}{2}),(m^{g}_i, \frac{1}{2}) \end{aligned}$\\
  \hline
$\alpha$-$\mu$ Fading \cite[Table IIV]{FoxH}:
$\begin{aligned} & f_{\mid y_i \mid}(x)=\frac{\sqrt{\eta^{y}_i}}{\Gamma(\mu^{y}_i)} {\rm H}_{0,1}^{1,0}\left[ \sqrt{\eta^{y}_i} x \left|
\begin{array}{ccc} - \\ (\mu^{y}_i-\frac{1}{\alpha^{y}_i},\frac{1}{\alpha^{y}_i}) \end{array}\right. \right]\end{aligned}$ \text{where} $\eta^{y}_i=\frac{\Gamma(\mu^{y}_i+\frac{2}{\alpha^{y}_i})}{\Gamma(\mu^{y}_i)}$
  &$\begin{aligned} \quad \quad & { \Pi}(\rho, N)=
\left(\prod_{i=1}^{{N}} \Gamma(\mu^{h}_i)\Gamma(\mu^{g}_i)\right)^{-1}\\ &{\rm H}_{0,1:1,2,\ldots, 1, 2}^{0,0:2, 1,\ldots, 2, 1}\left[\!\!\!\begin{array}{ccc} \sqrt{\eta^{h}_1 \eta^{g}_1 } \sqrt{\rho_t}\\ \vdots \\ \sqrt{\eta^{h}_N \eta^{g}_N } \sqrt{\rho_t}\end{array} \!\!\left |\!\!\begin{array}{ccc}  -: (1,1), -; \ldots; (1,1), - \\(0;1,\ldots, 1): \{\xi_{1},\Xi_{1}\}; \ldots;  \{\xi_{N},\Xi_{N}\} \end{array}\right. \!\!\!\!\right]\\ & \quad\quad \quad  \{\xi_{i},\Xi_{i}\}=(\mu^{h}_i, \frac{1}{\alpha^{h}_i}),(\mu^{g}_i, \frac{1}{\alpha^{g}_i}) \end{aligned}$\\ \hline
  Rice Fading \cite[Eq. (2.16)]{simon}
 $\begin{aligned} &  f_{\mid y_i \mid}(x)\!=\!\!\!\underset{K\longrightarrow\infty}{\lim}\sum_{k=0}^{K}\kappa^{y}_{ k} H_{0,1}^{1,0}\left[ c^{y}_{k} z \left|
\begin{array}{ccc} - \\ (k+\frac{1}{2},\frac{1}{2}) \end{array}\right. \right]\end{aligned}$  & $\begin{aligned}\quad \quad &{ \Pi}(\rho, N)\!\!\!=\!\!\!\sum_{k_1, \ldots, k_N=0}^{\infty}\sum_{t_1, \ldots, t_N=0}^{\infty}
\tau {\rm H}_{0,1:1, 2,\ldots, 1, 2}^{0,0:2, 1,\ldots,2, 1}\\&\Bigg[\!\!\!\begin{array}{ccc} \Delta \sqrt{\rho_t}\\ \vdots \\ \Delta \sqrt{\rho_t}\end{array} \!\!\Bigg| \begin{array}{ccc}  -: (1,1); \ldots; (1,1) \\(0;1,\ldots, 1)\!:\! (k_1\!+1\!,\!\frac{1}{2}), (t_1\!+\!1\!,\!\frac{1}{2}); \ldots;  (k_N\!+\!1\!,\! \frac{1}{2}),(t_N\!+\!1\!,\! \frac{1}{2})  \end{array}\Bigg. \!\!\!\!\Bigg]\end{aligned}$ \\ \hline
Fisher-Snedecor $\cal F$ \cite[Eq.(3)]{fisher}:
  $\begin{aligned} & f_{\mid y_i \mid}(x)=\frac{m^{y}_i}{m^{y}_{s_i} \Gamma(m^{y}_{s_i})\Gamma(m^{y}_i)} &\\& \quad \quad \times  {\rm H}_{1,1}^{1,1}\left[ \frac{m^{y}_i x}{m^{y}_{s_i} }\left|
\begin{array}{ccc} (-m^{y}_{s_i}+\frac{1}{2},\frac{1}{2}) \\ (m^{y}_i-\frac{1}{2},\frac{1}{2}) \end{array}\right. \right]\end{aligned}$
& $\begin{aligned} \quad \quad &{\Pi}(\rho, N)= \left(\prod_{i=1}^{{N}} \Gamma(m^{g}_{s_i})\Gamma(m^{g}_i)\Gamma(m^{h}_{s_i})\Gamma(m^{h}_i)\right)^{-1}
\\ &{\rm H}_{0,1:3,2,\ldots, 3, 2}^{0,0:2, 3,\ldots, 2, 3}\left[\!\!\!\begin{array}{ccc} \sqrt{\frac{m^{h}_1 m^{g}_1 }{m^{h}_{s_1}m^{g}_{s_1}} } \sqrt{\rho_t} \\ \vdots \\ \sqrt{\frac{m^{h}_N m^{g}_N }{m^{h}_{s_N}m^{g}_{s_N}} } \sqrt{\rho_t}\end{array} \!\!\left |\!\!\begin{array}{ccc}  -: (1,1), \{\delta_{1},\Delta_{1}\}; \ldots; (1,1), \{\delta_{N},\Delta_{N}\}\\(0;1,\ldots, 1): \{\xi_{1},\Xi_{1}\}; \ldots;  \{\xi_{N},\Xi_{N}\} \end{array}\right. \!\!\!\!\right]\\ & \quad\quad  \{\delta_{i},\Delta_{i}\}=(1-m^{h}_{si}, \frac{1}{2}),(1-m^{g}_{si}, \frac{1}{2}) \\& \quad \quad \{\xi_{i},\Xi_{i}\}=(m^{h}_{i}, \frac{1}{2}),(m^{g}_i, \frac{1}{2})\end{aligned}$ \\
\hline
Generalized ${\cal K}$  \cite[Table IIV]{FoxH}:
  $\begin{aligned} & f_{\mid y_i \mid}(x)=\frac{\sqrt{m^{y}_i k^{y}_i}}{ \Gamma(m^{y}_i)\Gamma(k^{y}_i)}& \\& \!\!\!\!\! \quad \times {\rm H}_{0,2}^{2,0}\left[x \sqrt{m^{y}_i k^{y}_i} \left|
\begin{array}{ccc} - \\ (m^{y}_i-\frac{1}{2},\frac{1}{2}), (k^{y}_i-\frac{1}{2},\frac{1}{2}) \end{array}\right. \right]\end{aligned}$
& $\begin{aligned} \quad \quad &{\Pi}(\rho, N)=
\left(\prod_{i=1}^{{N}} \Gamma(m^{h}_i)\Gamma(\kappa^{h}_i)\Gamma(m^{g}_i)\Gamma(\kappa^{g}_i)\right)^{-1}\\ &{\rm H}_{0,1:1,4,\ldots, 1, 4}^{0,0:4, 1,\ldots, 4, 1}\left[\!\!\!\!\!\begin{array}{ccc} \sqrt{m^{h}_1 \kappa^{h}_1 m^{g}_1 \kappa^{g}_1 } \sqrt{\rho_t}\\ \vdots \\ \sqrt{m^{h}_N \kappa^{h}_N m^{g}_N \kappa^{g}_N} \sqrt{\rho_t}\end{array} \!\!\!\!\left |\!\!\begin{array}{ccc}  -: (1,1), -; \ldots; (1,1), -\\(0;1,\ldots, 1): (\xi_{1},\Xi_{1})_{p_{11}}; \ldots;  (\xi_{N},\Xi_{N})_{p_{N}} \end{array}\right. \!\!\!\!\right]\\ & \quad\quad \quad  \{\xi_{i},\Xi_{i}\}=(m^{h}_i, \frac{1}{2}),(m^{g}_i, \frac{1}{2}),(k^{h}_i, \frac{1}{2}),(k^{g}_i, \frac{1}{2})\end{aligned}$ \\
\hline
\end{tabular}
\end{table*}

\subsection{Diversity Order and Coding Gain}
In this section, we analyze the diversity order and coding gain of RIS-assisted communications over generalized fading channels. In the current literature, the diversity order has been assessed by relying on approximations and bounds \cite{bour}-\!\!\cite{M4} (see Table I), and under Rayleigh fading. A common approach for analyzing the diversity order of RIS-aided systems is to leverage the CLT. However, this approach is accurate only for a large number of reconfigurable elements \cite{basar1}, and, in general, it is not sufficiently accurate for high-SNR analysis, which is the regime of interest for analyzing the diversity order \cite{CL5}. In \cite{bour}, for example, the diversity order was shown to be $\frac{N}{2}\frac{\pi^{2}}{16-\pi^{2}}$ in Rayleigh fading, which implies that the full diversity order cannot be obtained even in the absence of phase errors. By resorting to some bounds, however, the authors of \cite{CL5} recently showed that the full diversity order equal to $N$ is achievable in Rayleigh fading (even in the presence of phase errors). A detailed summary of the current methods and results on the diversity order of RIS-aided systems is available in Table I.

In what follows, building upon the high-SNR analysis of the exact outage probability in Proposition 1 and Proposition 2, we compute the exact diversity order and coding gain of RIS-assisted systems over generalized fading. We prove, in particular, that the diversity order in Rayleigh fading is $N$ and that it may exceed this value in less severe fading channels.

\vspace{0.25cm}
\textit{Proposition 3:}
Consider the multivariate Fox's H-function in  (\ref{pout}) and  define the set of  poles ${\cal S}=(\zeta_1,\ldots,\zeta_{N})$, where $\zeta_l=\underset{j=1,\ldots, \widetilde{m_l}}{\min} \left\{\frac{\xi_{l1}}{\Xi_{l1}}, \ldots, \frac{\xi_{l\widetilde{m}_l}}{\Xi_{l\widetilde{m}_l}}\right\}$. For each pole $s_j$, define the
set of indexes $K^{(l)}_j =\{k:k\in\{1,\ldots,\widetilde{m}_l\},r_{k,j}=\xi_{lj}-\Xi_{lj}\frac{\xi_{lk}}{\Xi_{lk}} \in\{0,1,2,\ldots\}\}$ and let $\widetilde{N}^{(l)}_{j_l} = |K^{(l)}_{j_l}|$ be
the multiplicity of the pole $\zeta_l$ with $j_l=\underset{j=1,\ldots, \widetilde{m_l}}{\arg \min}\left\{{\xi_{lj}}/{\Xi_{lj}}\right\}$.  The asymptotic expansion of (\ref{pout}) near $\rho_t=\frac{\gamma_{th}}{\rho}=0$ is \cite[Theorem 1.2]{kilbas}
\begin{eqnarray}
{ \Pi}(\rho, N)\approx\frac{\!\tau}{\Gamma(1+\sum_{l=1}^{N}\zeta_l)} \prod_{l=1}^{N} \!\widetilde{\Theta}(-\zeta_l)\left[\ln\left(\sqrt{\frac{\rho}{{\widetilde c}^{2}_l\gamma_{th}}}\right)\right]^{\widetilde{N}^{(l)}_{j_l}\!-\!1}{\widetilde c}_l^{\zeta_l}\left(\!\!\frac{\gamma_{th}}{\rho}\!\!\right)^{\!\!\frac{\zeta_l}{2}},
\label{asymoutr}
\end{eqnarray}
 where the constants $\widetilde{\Theta}_l$, $l=1,\ldots, N$, are given by
 \begin{eqnarray}
 \!\!\!\!\!\widetilde{\Theta}_l(\zeta_l)=\frac{1}{\Gamma\left(\widetilde{N}^{(l)}_{j_l}\right)}\left\{\prod_{k\in K^{(l)}_{j_l} }\frac{(-1)^{r_{k,j_l}}}{r_{k,j_l}!\Xi_k} \right\} \frac{\prod_{i\not\in K^{(l)}_{j_l}}\Gamma\left(\xi_{i}+\Xi_{i}\zeta_l\right)\prod_{i=1}^{{\widetilde n}_l}\Gamma\left(1-\delta_{i}-\Delta_{i}\zeta_l)\right)}{\prod_{i={\widetilde n}_l+1}^{{\widetilde p}_l}\Gamma\left(\delta_i+\Delta_{i}\zeta_l\right)\prod_{i={\widetilde m}_l+1}^{{\widetilde q}_l}\Gamma(1-\xi_{i}-\Xi_{i}\zeta_l)}.
 \label{thetai}
 \end{eqnarray}

\textit{Proof:}
Equation (\ref{asymoutr})  is obtained by evaluating the residues of the Mellin-Barnes integrals in (\ref{LMGF}) at the poles of the terms $\Gamma(\xi_{lj}+u_l\Xi_{lj})$, $j=1,\ldots, \widetilde{m_l}$, according to \cite[Theorem 1.2]{kilbas} and by keeping only the dominant terms using \cite[Eq. (1.8.3)]{kilbas}.

\vspace{0.25cm}
As mentioned, the Fox's H-function fading distribution generalizes many well-known
fading distributions, such as the  Rayleigh, Nakagami-$m$, and $\alpha$-$\mu$ distributions. It is interesting to analyze how the general
expressions derived for the asymptotic outage probability simplify when selecting the parameters corresponding to known distributions. We have three possible scenarios.
\begin{itemize}
    \item  Scenario 1: The  poles $\frac{\xi_{lj}}{\Xi_{lj}}$, $j=1,\ldots,\widetilde{m_l}$ are simple. This occurs when  $r_{k{j_l}}$ is neither a negative integer nor zero. In this case $\widetilde{N}^{(l)}_{j_l}=1$. This case study applies, for instance, to i.ni.d. $h_i$ and $g_i$ over Nakagami-$m$ and i.ni.d. $\alpha$-$\mu$, with $\alpha^{h}_i\mu^{h}_i\neq\alpha^{g}_i\mu^{g}_i$, fading channels.
\item Scenario 2: The poles  $\frac{\xi_{lj}}{\Xi_{lj}}$, $j=1,\ldots,\widetilde{m_l}$, all coincide. This occurs when  $\frac{\xi_{lj}}{\Xi_{lj}}=\frac{\xi_{lk}}{\Xi_{lk}}$, $k,j=1,\ldots,\widetilde{m_l}$, and in this case $\widetilde{N}^{(l)}_{j_l}=\widetilde{m}_l$. This case study includes, as special cases, Rayleigh, i.i.d. Nakagami-$m$, and $\alpha$-$\mu$, with $\alpha^{h}_i\mu^{h}_i=\alpha^{g}_i\mu^{g}_i$, fading channels, for which $\widetilde{N}^{(l)}_{j_l}=2$.
\item  Scenario 3: Some poles are simple and the others coincide, such that $r_{k{j_l}}=-r$, where $r$ is a positive integer. This case study, however, does not apply to (\ref{asymoutr}), since only the  smallest pole (the dominant pole), i.e.,  $j_l=\underset{j=1,\ldots, \widetilde{m_l}}{\arg \min}\left\{{\xi_{lj}}/{\Xi_{lj}}\right\}$, is considered in (\ref{asymoutr}).
\end{itemize}

\vspace{0.25cm}
Taking into consideration the just mentioned scenarios, the diversity order and the coding gain of RIS-assisted systems over generalized fading channels is stated in the following proposition.

\vspace{0.25cm}
\textit{Proposition 4:} Consider the general  Fox's-H fading model in (\ref{h}). The asymptotic (for high-SNR) outage probability of an RIS-aided system  can be formulated as
\begin{equation}
{ \Pi}(\rho, N)\underset{\rho\rightarrow\infty}{\approx}({\cal G}_c \rho )^{-{\cal G}_d},
\label{asymout}
\end{equation}
where ${\cal G}_d$ denotes the diversity order given by
\begin{equation}
{\cal G}_d=\frac{\sum_{i=1}^{N}\underset{j=1,\ldots, \widetilde{m_l}}{\min}\left\{\frac{\xi_{ij}}{\Xi_{ij}}\right\}}{2},
\label{Gdopt}
\end{equation}
and ${\cal G}_c$ denotes that coding gain given by
\begin{eqnarray}
{\cal G}_c= \left\{
  \begin{array}{ll}
   \!\!\! \frac{1}{\gamma_{th}}\left(\frac{\tau\left(\prod_{i=1}^{N} \widetilde{\Theta}_i(-\zeta_i){\widetilde c}_i^{\zeta_i}\right)}{\Gamma(1+\sum_{i=1}^{N}\zeta_i)}\right)^{-\frac{2}{\sum_{i=1}^{N}\zeta_i}} &\!\!\!\!\!\!\!\!\!\!\!\! \hbox{\!\!\!\!\!\!\! \quad \quad \, Scenario 1, } \\
    \!\!\!\!\frac{1}{\gamma_{th}}\left(\!\!\frac{\tau\left(\prod_{i=1}^{N} \widetilde{\Theta}_i(-\zeta_i)\ln\left(\frac{\rho }{{\widetilde c}^{2}_i\gamma_{th}}\right){\widetilde c}_i^{\zeta_i}\!\!\!\right)}{\Gamma(1+\sum_{i=1}^{N}\zeta_i)}\right)^{-\frac{2}{\sum_{i=1}^{N}\zeta_i}}  & \!\!\!\!\!\!\!\!\!\! \; \hbox{\!\!\!\!\!\!\! \quad \quad Scenario 2},
  \end{array}
\right.
\label{gcl}
\end{eqnarray}
 where $\zeta_l=\underset{j=1,\ldots, \widetilde{m_l}}{\min}\left\{{\xi_{lj}}/{\Xi_{lj}}\right\}$,
  and the constants
 $\widetilde{\Theta}_l(\zeta_l)$  are given in Proposition 3 with $\widetilde{N}^{(l)}_{j_l}=1$  and  $\widetilde{N}^{(l)}_{j_l}=2$ for Scenario 1 and 2, respectively.

\textit{Proof:} It follows from Proposition 3, under the assumptions stated in Scenarios 1 and 2.

\vspace{0.25cm}
\textit{Remark 3:} The usually considered Rayleigh fading channel model (see Table I) can be retrieved from Proposition 4 by considering Scenario 2 with $\zeta_l=2$ for $l=1,\ldots, N$.

\vspace{0.25cm}
\textit{Proposition 5:} Consider an RIS-assisted communication system over Rician fading. From (\ref{poutr}), the asymptotic outage probability is
\begin{eqnarray}
{ \Pi}(\rho, N)\approx \frac{2^{N}(1+K^{h})^{N}(1+K^{g})^{N}}{\Gamma(1+2N)} e^{-N( K^{h}+K^{g})} \ln\left(\frac{\rho}{\Delta^{2} \gamma_{th}}\right)^{N}\left(\frac{\rho}{\gamma_{th}}\right)^{-N}.
\label{outrice}
\end{eqnarray}

\textit{Proof:} It follows by computing the residues at $\zeta_l=\min\{2 k_{l}+2, 2 t_l+2\}, l=1, \ldots, N$, and by keeping only the dominant term of the infinite series expansion in (\ref{poutr}), which corresponds to $k_l=t_l=0$. Then, steps similar to Proposition 3 yield (\ref{outrice}) with the aid of some manipulations.

\vspace{0.25cm}
By substituting the specific parameters of the fading models summarized in Table II into the generalized expressions of the outage probability in Propositions  4  and 5 (under the assumptions of Scenarios 1 and 2), it is possible to obtain explicit expressions for the corresponding diversity order ${\cal G}_d$ and coding gain ${\cal G}_c$. These are reported in Table III, from which the following important conclusions and performance trends are unveiled.

\begin{itemize}
\item Over Rayleigh fading channels, the obtained results coincide with those derived in earlier research works as reported in Table I, but by using different methods of analysis, e.g., approximations and bounds.

  \item RIS-aided systems achieve a diversity order equal to $N$ for an arbitrary number $N$ of reconfigurable elements in both Rayleigh and Rice fading. The impact of the LoS component, i.e., $K$,  is mainly
reflected in the achievable coding gain. In particular, a strong LoS
component (large $K$) is beneficial.

\item Under severe fading $m^{h}_i, m^{g}_i<1$ or $\alpha^{h}_i\mu^{h}_i, \alpha^{g}_i\mu^{g}_i<1$, RIS-aided systems achieve a diversity order less than $N$. However, such a scenario may not occur in optimized deployments in which the RISs are positioned in order to leverage the LoS paths with the transmitter and possibly with the receiver. In these cases, $m^{h}_i$ and $m^{g}_i$ are, in fact, relatively large.

\item Under fading channels less severe than Rayleigh, e.g., $m^{h}_i$ and $m^{g}_i$ are larger than one, a diversity order greater than $N$ can be obtained.

\item Under the assumption of Scenario 2, which encompasses i.i.d. Rayleigh fading, the scaling law of the outage probability as a function of $\rho \to \infty$ is ${\left( {{{\ln \left( \rho  \right)} \mathord{\left/ {\vphantom {{\ln \left( \rho  \right)} \rho }} \right. \kern-\nulldelimiterspace} \rho }} \right)^N}$. This trend is in agreement with \cite{CL2}, where it was proved by using exact analysis for $N=1$ and upper and lower bounds for $N > 1$. Similar trends were reported in \cite{CL5}. It is worth noting that this scaling law holds true for Rice fading as well, as unveiled, for the first time in the literature, by \eqref{outrice}. As remarked in \cite{CL2}, this is a new scaling law, which generalizes the definitions of diversity order and coding gain typically used in wireless communications \cite{gian}.

\item Under the assumptions of Scenario 1 (e.g., i.ni.d. Nakagami-m and $\alpha$-$\mu$ fading in Table III), the scaling law ${\left( {{{\ln \left( \rho  \right)} \mathord{\left/ {\vphantom {{\ln \left( \rho  \right)} \rho }} \right. \kern-\nulldelimiterspace} \rho }} \right)^N}$ does not emerge, and the outage probability scales as ${\rho ^{ - N}}$ for $\rho \to \infty$. To be best of the authors knowledge, this difference in the scaling law between Scenario 1 and Scenarios 2 was never reported in the literature.

\end{itemize}

\vspace{0.25cm}
To the best of our knowledge, the analytical framework introduced in this section is the first one that is exact and is applicable to generalized fading channels. Also, the corresponding diversity analysis is not based either on approximations or bounds.

\begin{table*}
\caption{DIVERSITY AND CODING GAIN OF RIS-ASSISTED SYSTEMS OVER
WIDELY USED FADING CHANNELS} 
\centering
\begin{tabular}{c|c|c}
  \hline
  \textbf{Distribution} & \textbf{Diversity Order} & \textbf{Coding Gain}  \\ \hline \hline
  Rayleigh & $N$ & $\frac{1}{\gamma_{th}}\left(-\ln\left(\frac{\gamma_{th}}{\rho}\right)^{N}\frac{2^{N}}{\Gamma(1+2N)}\right)^{-\frac{1}{N}}$ \\ \hline
  i.ni.d. Nak.-$m$ & $\sum
  \limits_{i=1}^{N}\min\{m^{h}_i, m^{g}_i\}$& $\!\!\!\!\frac{1}{\gamma_{th}}\!\!\left(\!\!\frac{2^{N}\prod_{j=1}^{N}(m^{h}_jm^{g}_j)^{\min\{m^{h}_j, m^{g}_j\}}\Gamma\left(m^{{\underset{x\in\{h,g\}}{x\neq\arg\min} \{m^{x}_j\}}}_j-\min\{m^{h}_j, m^{g}_j\}\right)}{ \left(\prod_{i=1}^{{N}} \Gamma(m^{h}_i)\Gamma(m^{g}_i)\right)\Gamma\left(1+2\sum_{i=1}^{N}\min\{m^{h}_i, m^{g}_i\}\right)}\!\!\right)^{\frac{1}{\sum\limits_{i=1}^{N}\min\{m^{h}_i, m^{g}_i\}}}$ \\ \hline
  i.ni.d. $ \alpha$-$\mu$ \\$\alpha^{h}_i\mu^{h}_i\neq\alpha^{g}_i\mu^{g}_i$& $\frac{\sum\limits_{i=1}^{N}\min\{\alpha^{h}_i \mu^{h}_i, \alpha^{g}_i \mu^{g}_i\}}{2}$ &\!\!\!\!\! $\frac{1}{\gamma_{th}}\!\!\left(\!\!\frac{\prod_{j=1}^{N}\alpha^{\underset{x\in\{h,g\}}{c=\arg\min} \{\alpha^{x}_j\mu^{x}_j\}}_j\Gamma\left(\mu^{x\in\{h,g\}\neq c}_j-\frac{1}{\alpha^{x\in\{h,g\}\neq c}_j}\min\{\alpha^{h}_j\mu^{h}_j, \alpha^{g}_j \mu^{g}_j\}\right)}{ \prod_{j=1}^{N}\left(\left(\sqrt{\eta^{h}_j\eta^{g}_j}\right)^{-\min\{\alpha^{h}_j\mu^{h}_j,\alpha^{g}_j\mu^{g}_j\}} \Gamma(\mu^{h}_j)\Gamma(\mu^{g}_j)\right)\Gamma\left(1+\sum_{i=1}^{N}\min\{\alpha^{h}_i\mu^{h}_i, \alpha^{g}_i \mu^{g}_i\}\right)}\!\!\right)^{\!\!\!\frac{2}{\sum\limits_{i=1}^{N}\min\{\alpha^{h}_i \mu^{h}_i, \alpha^{g}_i \mu^{g}_i\}}}$\\ \hline
Rice & $N$ & $\frac{1}{\gamma_{th}}\frac{e^{( K^{h}+K^{g})}\left(-\ln\left(\frac{\Delta^{2}\gamma_{th}}{\rho}\right)^{-1}\right)\Gamma(1+2N)^{\frac{1}{N}}}{2(1+K^{h})(1+K^{g})}$\\
   \hline
\end{tabular} 
\end{table*}

\section{Outage Probability - With Phase Noise}
In practice, the phase shifts of the reconfigurable elements of an RIS cannot be optimized with an arbitrary precision, e.g., because of the finite number of quantization bits used or because of errors when estimating the phases of the fading channels \cite{CL4}, \cite{renzoquant1}. In these cases, the phase of the $i$th element of the RIS can be written as $\phi_i=-\angle{h_i}-\angle{g_i}+\theta_i$, where $\theta_i$ denotes a random phase noise, which is assumed to be i.i.d. in this paper. Thus, the equivalent channel observed by the  receiver is a complex random variable and the SNR is
\begin{equation}
\gamma=\rho\left|\sum_{i=1}^{N} |h_i| |g_i| e^{j \theta_i}\right|^{2}=\rho|H|^{2}.
\label{snoise}
\end{equation}

We assume that the distribution of $\theta_i$ is arbitrary but its mean is zero. Examples of phase noise distributions include Gaussian, generalized uniform, and uniform RVs whose characteristic functions, $\mathbb{E}\{e^{j t \theta}\}={\cal K}_t$, are as follows:
\begin{enumerate}
  \item Gaussian $\theta_i\sim {\cal N}(0, \sigma^{2})$ \cite{th} \begin{equation}{\cal K}_t\approx e^{-\sigma^{2}\frac{ t^{2}}{2}},
\end{equation}
  \item Generalized uniform $\theta_i\sim {\cal U}(-q\pi, q\pi)$, $q\neq1$, \cite{renzoquant1}
\begin{equation}{\cal K}_t=  \frac{\sin(q\pi t)}{q\pi t},
\end{equation}
\item Uniform $\theta_i\sim {\cal U}(-\pi, \pi)$ \cite{renzoquant1}
\begin{equation}{\cal K}_t= \frac{\sin(\pi t)}{\pi t},
\end{equation}
\end{enumerate}
where $(a)$ follows from $\mathbb{E}\{e^{j t \theta_i}\}_{\theta_i  \in [\pi,\pi]}\underset{\sigma^{2}\ll 1}{\approx} \mathbb{E}\{e^{j t \theta_i}\}_{\theta_i \in [-\infty,\infty]}$, which
stems from the fact that, in
practice, we are interested in standard deviations of only a few
degrees.

It is worth noting that the Gaussian distribution is versatile to
represent continuous phase errors. In \cite{th}, e.g., it was shown that the phase errors are Gaussian distributed under widely applicable assumptions. Likewise, due to hardware limitations, only a finite number of phase shifts can be realized, which leads to quantization errors. In this case, the generalized uniform distribution constitutes a versatile model to account for the quantization noise by setting $q=2^{-L}$, where $L\geq1$ is a positive
integer that denotes the number of quantization bits used \cite{CL4}, \cite{renzoquant1}.

The SNR in (\ref{snoise}) is formulated in terms of the square magnitude of a linear combination of complex random variables with random magnitudes
and random phases. In general, the calculation of the exact distribution of the SNR in (\ref{snoise}) is an open research issue, and is very intricate for arbitrary values of $N$. To tackle this issue, we proceed as follows: (i) first, we study the distribution of the SNR in (\ref{snoise}) under the assumption of a large number of reconfigurable elements $N$ of the RIS, i.e., $N \gg 1$. The obtained analytical framework is based on the CLT and is typically appropriate for analyzing RIS-aided systems with practical numbers of reconfigurable elements and for typical values of $\rho$. As noted in, e.g., \cite{CL5}, the resulting analysis is usually not accurate in the high-SNR regime (i.e., for $\rho \to \infty$), and, therefore, for analyzing the attainable diversity order; and (ii) then, we introduce upper and lower bounds for the SNR in (\ref{snoise}) in the presence of phase noise. The objective is to identify sufficient conditions for achieving the full diversity order. The main peculiarity of the latter approach lies in its applicability to RIS-aided systems with an arbitrary number of reconfigurable elements $N$ and for any SNR regime.

\subsection{Performance Analysis for Large $N$ and Finite Values of $\rho$}

In this sub-section, we introduce an analytical framework for computing the outage probability of RIS-aided systems in the presence of phase noise and under the assumption $N \gg 1$.

\vspace{0.25cm}
\textit{Proposition 6:} Define $\Lambda_1=\mathbb{E}\{|h_i|\}$ and $\Lambda_2=\mathbb{E}\{|g_i|\}$. For large values of $N$, the outage probability in the presence of phase noise can be formulated as follows.
\begin{itemize}
  \item Case 1: In the presence of Gaussian and generalized uniform phase noise, we have
\end{itemize}
\begin{eqnarray}
{ \Pi}(\rho, N)\!=\!\frac{1} {2\pi}{\rm H}_{3,1:1,1;2,1; 2, 2}^{0,2:1,1;1, 1; 2,1}\Bigg[\!\!\!\begin{array}{ccc} \frac{-\gamma_{th}}{N^{2}4\nu^{2}\rho}\\ \frac{\gamma_{th}}{N^{2} \nu^{2}\rho}\\ \frac{\gamma_{th}}{N^{2}\nu^{2}\rho}\end{array} \!\!\Bigg|\begin{array}{ccc} {\cal B}; (0,2); (1,1),\left(\frac{\nu^{2}}{2 \sigma_X^{2}},2\right);(1,1),\left(\frac{\nu^{2}}{2 \sigma_Y^{2}},2\right)\\(1;0,-1,-1):(0,1);\left(\frac{\nu^{2}}{2 \sigma_X^{2}},1\right);\left( \frac{\nu^{2}}{2 \sigma_Y^{2}},1\right),\left(\frac{1}{2},1\right) \end{array}\Bigg. \Bigg],
\label{outsig}
\end{eqnarray}
where $  {\cal B}=\{(1;-1,-1,-1), \big(\frac{1}{2};-1,0,-1\big), (1;0,-1,-1)\}$, $\nu={\cal K}_1 \Lambda_1\Lambda_2$, $\sigma_Y^{2}=\frac{1}{2N}\left(1-{\cal K}_2\right)$, and $\sigma^{2}_X=\frac{1}{2N}\left(1+ {\cal K}_2-2{\cal K}_1^{2}\Lambda_1^{2}\Lambda_2^{2}\right)$.
\begin{itemize}
       \item Case 2: In the presence of uniform phase noise, we have
     \end{itemize}
 \begin{eqnarray}
{\Pi}(\rho, N)=\pi^{-1} {\rm H}_{0,1:2,2; 2, 2}^{0,0:2, 1; 2, 1}\Bigg[\begin{array}{ccc}\frac{\gamma_{th}}{N^{2}\rho} \\ \frac{\gamma_{th}}{N^{2}\rho}\end{array} \Bigg|  \begin{array}{ccc} -: \left(N,2\right), (1,1);\left(N,2\right),(1,1)\\ (0;1,1)\!:\!\left(N,1\right),\left(\frac{1}{2},1\right); \left(N,1\right),\left(\frac{1}{2},1\right) \end{array}\Bigg. \Bigg].
\label{outun}
\end{eqnarray}

\textit{Proof:} The proof is based on the application of the CLT. See Appendix A.

\vspace{0.25cm}
To the best of our knowledge, Proposition 6 is a new result that is not available in the literature and is applicable to generalized fading distributions in the presence of phase noise.

\subsection{Diversity Analysis for Large $N$}

Based on Proposition 6, this sub-section studies the outage probability in the high-SNR regime, i.e., for $\rho \to \infty$. It is known, however, that the CLT may not be suitable for analyzing the outage probability for $\rho \to \infty$ if $N$ is fixed. The analysis of this sub-section serves, therefore, as a benchmark for better understanding the limitations of the CLT when applied to RIS-aided systems for analyzing their performance in the high-SNR regime. This is elaborated next.

\vspace{0.25cm}
\textit{Proposition 7:}
Assume $\rho \to \infty$. Based on Proposition 6, the asymptotic outage probability in the presence of phase noise can be formulated  as
\begin{eqnarray}
{ \Pi}(\rho, N)\underset{\rho\rightarrow\infty}{\approx} \left\{
                                                                                       \begin{array}{ll}
                                                                                        {\cal A}_G \left(\frac{\rho}{\gamma_{th}}\right)^{-\left(\frac{\nu^{2}}{2\sigma^{2}_X}+\frac{1}{2}\right)} & \hbox{Case 1}, \\

\frac{\Gamma\left(N-\frac{1}{2}\right)^{2}}{\Gamma\left(N\right)^{2}}\left(\frac{\rho}{\gamma_{th}}\right)^{-1} & \hbox{Case 2},
                                                                                       \end{array}
                                                                                     \right.
\label{diph1}
\end{eqnarray}
 where
\begin{eqnarray}
\!\!\!\!\!\!\!\!\!\!\!\!{\cal A}_G &=& \frac{\left(N^{2} \nu^{2}\right)^{\!-\left(\frac{\nu^{2}}{2\sigma^{2}_X}\!+\!\frac{1}{2}\right)}}{\sqrt{\pi}}\sqrt{\frac{\nu^{2}}{2\sigma^{2}_Y}}
 \frac{\Gamma\left(\frac{\nu^{2}}{2\sigma^{2}_X}\right)}{\Gamma\left(\!-\frac{\nu^{2}}{2\sigma^{2}_X}\!\right)}\frac{\Gamma\left(\frac{1}{2}\!-\!\frac{\nu^{2}}{2\sigma^{2}_X}\right)}
{\Gamma\left(\!\frac{3}{2}\!+\!\frac{\nu^{2}}{2\sigma^{2}_X}\!\right)}.
\label{ag}
\end{eqnarray}

\textit{Proof:} Equation (\ref{diph1}) is obtained by computing the residues of
the integrand in (\ref{perint1}) and (\ref{per22}) by using \cite[Eqs. (1.84,1.85)]{kilbas}. As for Case 1, the residue is computed at the
points $ u_1= \frac{\nu^{2}}{2\sigma^{2}_X}$,  $u_2= \min\left\{\frac{\nu^{2}}{2\sigma^{2}_Y}, \frac{1}{2}\right\}=\frac{1}{2}$, and $u_3=0$.  As for  Case 2, the residue is computed at  $(u_1, u_2)=(\frac{1}{2},\frac{1}{2})$.

\vspace{0.25cm}
From (\ref{diph1}), we evince that the diversity order based on the CLT approximation for $N \gg 1$ is
\begin{eqnarray}
{\cal G}^{\rm{CLT}}_d &=& \left\{
                                                                                    \begin{array}{ll}
                                                                                        N {\cal E}+\frac{1}{2} & \hbox{Case 1}, \\\\
                                                                1 & \hbox{Case 2},
                                                                                       \end{array}
                                                                                     \right.
                                                                                     \label{dphase1}
\end{eqnarray}
where ${\cal E}=({\cal K}_1 \Lambda_1\Lambda_2)^2/\left(1+ {\cal K}_2-2{\cal K}_1^{2}\Lambda_1^{2}\Lambda_2^{2}\right)$.

\vspace{0.25cm}
\textit{Remark 4:} Based on (\ref{dphase1}), we conclude that, in general, the diversity order in the presence of phase noise is less that the full diversity order that is achievable in the absence of phase noise, which is given in (\ref{Gdopt}). As a case study, let us consider that the phase noise originates from the quantization bits $L$ used for the phase shifts. From (\ref{dphase1}), we obtain ${\cal E}=\frac{1}{2}$ for $L=1$ and ${\cal E}<1$ for $L>1$. Over Rayleigh fading, in particular, we obtain ${\cal G}^{\rm{CLT}}_d \approx 0.78N+0.5$ if $L=2$ and  ${\cal G}^{\rm{CLT}}_d \approx 0.8N+0.5$ if $L\rightarrow\infty$ (i.e., no phase noise). Therefore, we evince that  ${\cal G}^{\rm{CLT}}_d<{\cal G}_d$ for finite and infinite values of $L$, which is in disagreement with (\ref{Gdopt}) in the absence of phase noise (i.e., $L\rightarrow\infty$).

\vspace{0.25cm}
The example in Remark 4 confirms the unsuitability of the CLT for high-SNR analysis, and, in particular, for estimating the diversity order of RIS-assisted systems. To further corroborate the statements in Remark 4, let us assume $N=1$ in \eqref{snoise}. In this case, we would have $\gamma \left( {N = 1} \right) = \rho {\left( {\left| {{h_1}} \right|\left| {{g_1}} \right|\cos \left( {{\theta _1}} \right)} \right)^2} + \rho {\left( {\left| {{h_1}} \right|\left| {{g_1}} \right|\sin \left( {{\theta _1}} \right)} \right)^2} = \rho {\left| {{h_1}} \right|^2}{\left| {{g_1}} \right|^2}$, which implies that the SNR is independent of the phase error. This is different from (\ref{dphase1}).

In the next sub-section, we introduce sufficient conditions for ensuring that the full diversity order is achieved in RIS-assisted communications impaired by phase noise.

\subsection{Diversity Analysis for Arbitrary $N$}

The analytical framework introduced in the previous section based on the CLT is usually accurate for analyzing the performance of RIS-assisted communications for practical values of $N$ and for typical values of the SNR. However, it is not sufficiently accurate for estimating the diversity order, i.e., for $\rho \to \infty$. In the present paper, for these reasons, we do not attempt to introduce approximated analytical frameworks but focus our attention on identifying sufficient conditions for guaranteeing that RIS-aided systems achieve the full diversity order even in the presence of phase noise. This is a fundamental open issue for designing and optimizing RIS-aided systems. For example, the approach introduced in this section allows us to identify the minimum number of quantization bits that are needed for ensuring no diversity loss. This specific problem has been recently analyzed in \cite{CL5}, where it is shown that, under i.i.d. Rayleigh fading, two quantization bits (i.e., $L=2$) are necessary. The approach proposed in \cite{CL5} is specifically tailored for analyzing the impact of quantization bits in the presence of  i.i.d. Rayleigh fading. The analytical approach proposed in this section, on the other hand, is applicable to arbitrary distributions for the channel fading and for the phase noise.

To this end, we re-write the end-to-end SNR in \eqref{snoise} as $\gamma  = \rho {\left| H \right|^2} = \rho \left( {{X^2} + {Y^2}} \right)$, where $X=\sum_{i=1}^{N}|h_i| |g_i|\cos(\theta_i)$ and $Y=\sum_{i=1}^{N}|h_i| |g_i|\sin(\theta_i)$. Also, we define ${\varepsilon _{\min }} = \hspace{-0.25cm} \mathop {\min }\limits_{n \ne m \in \left[ {1,N} \right]} \left\{ {\cos \left( {{\theta _n} - {\theta _m}} \right)} \right\}$. The main results are stated in the following two propositions.

\vspace{0.25cm}
\textit{Proposition 8:} Assume ${\varepsilon _{\min }} = \hspace{-0.25cm} \mathop {\min }\limits_{n \ne m \in \left[ {1,N} \right]} \left\{ {\cos \left( {{\theta _n} - {\theta _m}} \right)} \right\} \ge 0$. The SNR in \eqref{snoise} is upper and lower bounded as follows
\begin{eqnarray}
\rho \sum\limits_{i = 1}^N {{{\left| {{h_i}} \right|}^2}{{\left| {{g_i}} \right|}^2}}  \le \rho \left( {{X^2} + {Y^2}} \right) \le \rho {\left( {\sum\limits_{i = 1}^N {\left| {{h_i}} \right|\left| {{g_i}} \right|} } \right)^2}
\end{eqnarray}

\textit{Proof:} Be definition, we have
\begin{eqnarray}
{X^2} = \sum\limits_{n = 1}^N {{{\left| {{h_n}} \right|}^2}{{\left| {{g_n}} \right|}^2}{{\cos }^2}\left( {{\theta _n}} \right)}  + \sum\limits_{n = 1}^N {\sum\limits_{m \ne n = 1}^N {\left( {\left| {{h_n}} \right|\left| {{g_n}} \right|\cos \left( {{\theta _n}} \right)} \right)\left( {\left| {{h_m}} \right|\left| {{g_m}} \right|\cos \left( {{\theta _m}} \right)} \right)} }
\end{eqnarray}
\begin{eqnarray}
{Y^2} = \sum\limits_{n = 1}^N {{{\left| {{h_n}} \right|}^2}{{\left| {{g_n}} \right|}^2}{{\sin }^2}\left( {{\theta _n}} \right)}  + \sum\limits_{n = 1}^N {\sum\limits_{m \ne n = 1}^N {\left( {\left| {{h_n}} \right|\left| {{g_n}} \right|\sin \left( {{\theta _n}} \right)} \right)\left( {\left| {{h_m}} \right|\left| {{g_m}} \right|\sin \left( {{\theta _m}} \right)} \right)} }
\end{eqnarray}

By using the identity $\cos \left( {\alpha  + \beta } \right) = \cos \left( \alpha  \right)\cos \left( \beta  \right) - \sin \left( \alpha  \right)\sin \left( \beta  \right)$, we obtain
\begin{eqnarray} \label{LowerBound}
{X^2} + {Y^2} &=& \sum\limits_{n = 1}^N {{{\left| {{h_n}} \right|}^2}{{\left| {{g_n}} \right|}^2}}  + \sum\limits_{n = 1}^N {\sum\limits_{m \ne n = 1}^N {\left| {{h_n}} \right|\left| {{g_n}} \right|\left| {{h_m}} \right|\left| {{g_m}} \right|\cos \left( {{\theta _n} - {\theta _m}} \right)} } \nonumber \\
& \mathop  \ge \limits^{\left( a \right)}& \sum\limits_{n = 1}^N {{{\left| {{h_n}} \right|}^2}{{\left| {{g_n}} \right|}^2}}  + {\varepsilon _{\min }}\sum\limits_{n = 1}^N {\sum\limits_{m \ne n = 1}^N {\left| {{h_n}} \right|\left| {{g_n}} \right|\left| {{h_m}} \right|\left| {{g_m}} \right|} } \nonumber \\
& \mathop  \ge \limits^{\left( b \right)}&\sum\limits_{n = 1}^N {{{\left| {{h_n}} \right|}^2}{{\left| {{g_n}} \right|}^2}}
\end{eqnarray}
\begin{eqnarray} \label{UpperBound}
{X^2} + {Y^2} &=& \sum\limits_{n = 1}^N {{{\left| {{h_n}} \right|}^2}{{\left| {{g_n}} \right|}^2}}  + \sum\limits_{n = 1}^N {\sum\limits_{m \ne n = 1}^N {\left| {{h_n}} \right|\left| {{g_n}} \right|\left| {{h_m}} \right|\left| {{g_m}} \right|\cos \left( {{\theta _n} - {\theta _m}} \right)} } \nonumber \\
&\mathop  \le \limits^{\left( c \right)}& \sum\limits_{n = 1}^N {{{\left| {{h_n}} \right|}^2}{{\left| {{g_n}} \right|}^2}}  + \sum\limits_{n = 1}^N {\sum\limits_{m \ne n = 1}^N {\left| {{h_n}} \right|\left| {{g_n}} \right|\left| {{h_m}} \right|\left| {{g_m}} \right|} } \nonumber \\
& =& {\left( {\sum\limits_{n = 1}^N {\left| {{h_n}} \right|\left| {{g_n}} \right|} } \right)^2}
\end{eqnarray}
where (a) and (b) follow under the assumption ${\varepsilon _{\min }} \ge 0$, and (c) follows because $\cos \left( {{\theta _n} - {\theta _m}} \right) \le 1$ for any phase errors. This concludes the proof.

\vspace{0.25cm}
\textit{Proposition 9:} If ${\varepsilon _{\min }} = \mathop {\min }\limits_{n \ne m \in \left[ {1,N} \right]} \left\{ {\cos \left( {{\theta _n} - {\theta _m}} \right)} \right\} \ge 0$, RIS-assisted transmission achieves the full diversity order in the presence of phase noise.

\textit{Proof:} It follows from Proposition 8, by noting the following: (i) the upper bound in \eqref{UpperBound} coincides with the SNR in the absence of phase errors, which is shown to achieve the full diversity order in Section III, and (ii) the lower bound in \eqref{LowerBound} is the SNR of an equivalent maximal-ratio combining system whose links have an SNR equal to ${{{\left| {{h_n}} \right|}^2}{{\left| {{g_n}} \right|}^2}}$. From \cite{gian}, the diversity order of the lower bounds in \eqref{LowerBound} is the same as the diversity order of the upper bound in \eqref{UpperBound}, since the latter bound corresponds to the SNR (scaled by a fixed constant) of an equivalent equal-gain combining system. This concludes the proof.

\vspace{0.25cm}
\textit{Remark 5:} The upper bound in \eqref{UpperBound} can be applied only if ${\varepsilon _{\min }} = \mathop {\min }\limits_{n \ne m \in \left[ {1,N} \right]} \left\{ {\cos \left( {{\theta _n} - {\theta _m}} \right)} \right\} \ge 0$. This implies that Proposition 9 yields a sufficient condition for achieving the full diversity order. If ${\varepsilon _{\min }} = \mathop {\min }\limits_{n \ne m \in \left[ {1,N} \right]} \left\{ {\cos \left( {{\theta _n} - {\theta _m}} \right)} \right\} < 0$, in other words, a diversity loss may occur.

\vspace{0.25cm}
Based on Proposition 8 and Proposition 9, the following remarks can be made:

\begin{itemize}
\item The condition ${\varepsilon _{\min }} = \mathop {\min }\limits_{n \ne m \in \left[ {1,N} \right]} \left\{ {\cos \left( {{\theta _n} - {\theta _m}} \right)} \right\} \ge 0$ implies that the absolute difference between pairs of phase errors is always less than $\pi/2$. This yields important guidelines to make the design of RISs robust to the phase noise.

\item Assume that the phase noise is determined by the number $L$ of quantization bits used. Then, by definition,  we have $\cos \left( {{{2\pi } \mathord{\left/ {\vphantom {{2\pi } {{2^L}}}} \right. \kern-\nulldelimiterspace} {{2^L}}}} \right) \le \cos \left( {{\theta _n} - {\theta _m}} \right) \le 1$. This yields ${\varepsilon _{\min }} = \cos \left( \pi  \right) =  - 1$ for $L=1$, ${\varepsilon _{\min }} = \cos \left( {{\pi  \mathord{\left/ {\vphantom {\pi  2}} \right. \kern-\nulldelimiterspace} 2}} \right) = 0$ for $L=2$, and ${\varepsilon _{\min }}  > 0$ for $L>2$. Therefore, the full diversity order can be ensured if at least two quantization bits are used. This result is in agreement with \cite{CL5}, but generalizes it to arbitrary fading distributions and phase noise distributions.

\item The potential loss of diversity for ${\varepsilon _{\min }} = \mathop {\min }\limits_{n \ne m \in \left[ {1,N} \right]} \left\{ {\cos \left( {{\theta _n} - {\theta _m}} \right)} \right\} <0$ can be understood by considering the case study for $N=2$ and $L=1$ in Proposition 8. In this case, we obtain ${\left. {{X^2} + {Y^2}} \right|_{{\rm{worst\; case}}}} = {\left( {\left| {{h_1}} \right|\left| {{g_1}} \right| - \left| {{h_2}} \right|\left| {{g_2}} \right|} \right)^2}$. The negative sign in the latter equation is responsible for the potential loss of diversity order for ${\varepsilon _{\min }} <0$.
\end{itemize}

\vspace{0.25cm}
In conclusion, with the aid of the sufficient condition identified in Proposition 9, an RIS can be appropriately optimized in order to guarantee that the full diversity order is achieved. To the best of our knowledge, this result for arbitrary fading and phase noise distributions was never reported in the literature. A similar approach could be applied to the analysis of hardware impairments different from the phase noise.

\section{Numerical Results}

In this section, we report some numerical results in order to substantiate the obtained analytical expressions of the outage probability and the analysis of the diversity order and coding gain with the aid of Monte Carlo simulations. Unless otherwise stated, the SNR threshold is set to $\gamma_{th}=0$ dB. It is worth mentioning that the numerical results consider relatively small values of $N$ in order to better highlight the impact of the diversity order, which is the main focus of the present paper, similar to \cite{CL5}.

Figure~1 shows the outage probability of an RIS-assisted system in the absence of phase noise as a function of the average SNR (i.e., $\rho$), for several values of $N$ and under Nakagami-m fading for $m_i^{g}>m^{h}_i$, $i=1,\ldots,N$, with $\underset{i=1,\ldots,N}{\min}\{m^{h}_i\}=0.5$.  We observe that the exact expression of the outage probability in (\ref{pout}) and its corresponding high-SNR approximation in (\ref{asymout}) are in close agreement with Monte Carlo simulations. In particular, Fig.~1 confirms that correctness of the diversity analysis in Section III-A. As expected, the outage probability decreases significantly as the number $N$ of reconfigurbale elements of the RIS increases.
\begin{figure}[!t]
\centering
\includegraphics[scale=0.5]{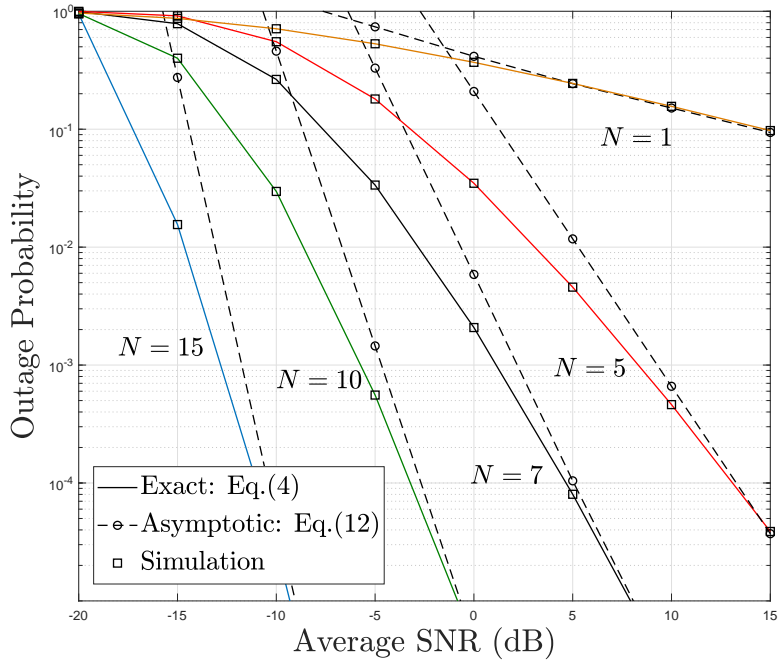}
\caption{Outage probability vs. the average SNR in Nakagami-$m$ fading.}
\end{figure}

Figure~2 shows the outage probability vs. the average SNR $\rho$ over $\alpha$-$\mu$ fading in both i.i.d and i.ni.d. scenarios. The conclusions are similar to those in Fig.~1. We observe, in particular, that the outage probability over i.ni.d. $\alpha$-$\mu$ fading decreases at a rate of $\rho^{-\sum_{i=1}^{N}\frac{\min\{\alpha^{h}_i \mu^{h}_i,\alpha^{g}_i \mu^{g}_i \}}{2}}$, in agreement with (\ref{asymoutr}) and more precisely with (\ref{asymout}) and (\ref{gcl}), under the assumptions of Scenario 1. Over i.i.d. fading, in addition, the figure confirms that the outage probability scales with ${{\ln \left( \rho  \right)} \mathord{\left/ {\vphantom {{\ln \left( \rho  \right)} \rho }} \right. \kern-\nulldelimiterspace} \rho }$, as predicted by (\ref{gcl}), under the assumptions of Scenario 2, and unveiled in \cite{CL2} and \cite{CL5} over i.i.d. Rayleigh fading channels.
\begin{figure}[!t]
\centering
\includegraphics[scale=0.5]{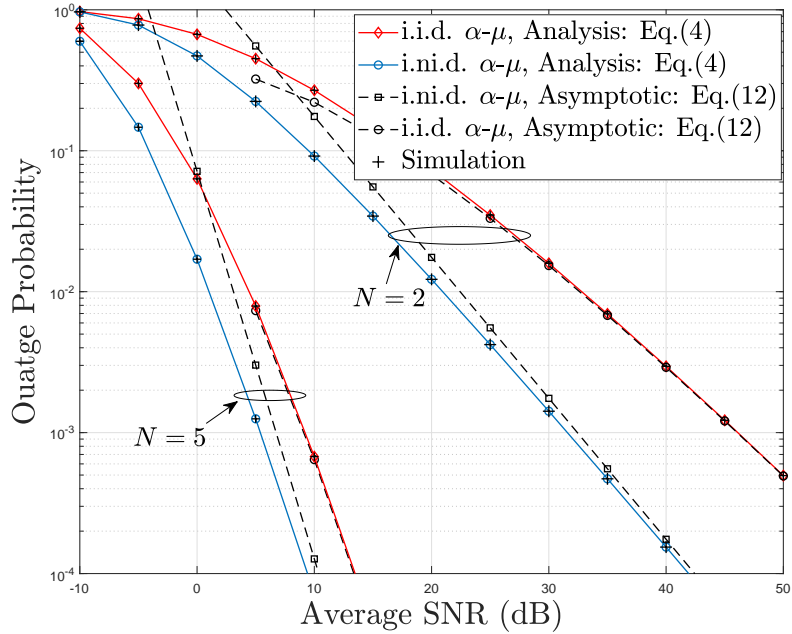}
\caption{Outage probability vs. the average SNR in $\alpha$-$\mu$ fading.}
\end{figure}

Figure 3 illustrates the outage probability of an RIS-aided system in the presence of phase noise over Nakagami-$m$ fading. The phase noise is modeled by assuming that the phases are quantized by using $L=1$ and $L=2$ quantization bits. The numerical results obtained with Monte Carlo simulations are in agreement with the analytical findings in Section IV. The figure confirms, in particular, that the CLT does not yield, in general, an accurate estimate of the diversity order. In addition, we observe, that a two-bit quantization ($L=2$) for the phase shifts yields, in the considered case study, sufficiently good performance in terms of outage probability. In particular, the full diversity order can be achieved, as stated in Proposition 9.
\begin{figure}[!t]
\centering
\includegraphics[scale=0.5]{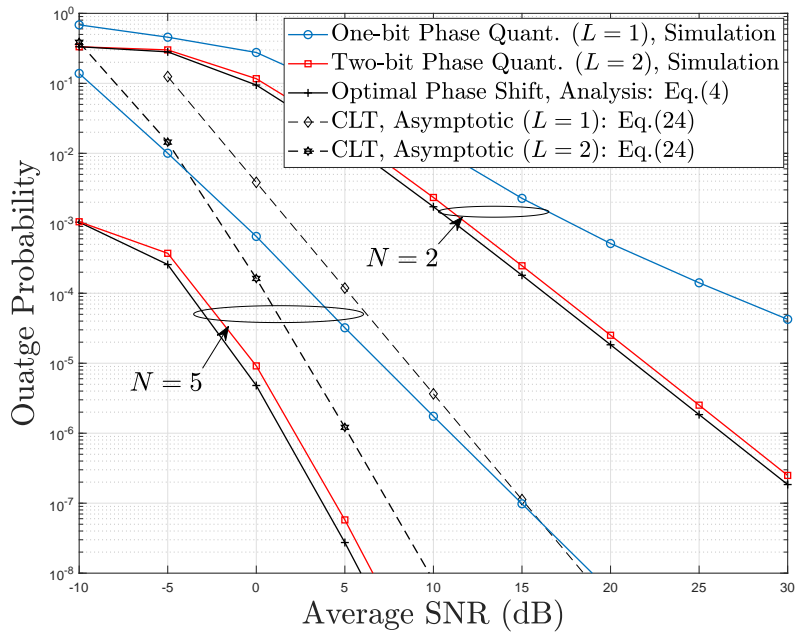}
\caption{Outage probability vs. the average SNR, for different values of $L$ and $N$ (Nakagami-$m$ fading with $m=1.5$).}
\end{figure}

\begin{figure}[!t]
\centering
\includegraphics[scale=0.5]{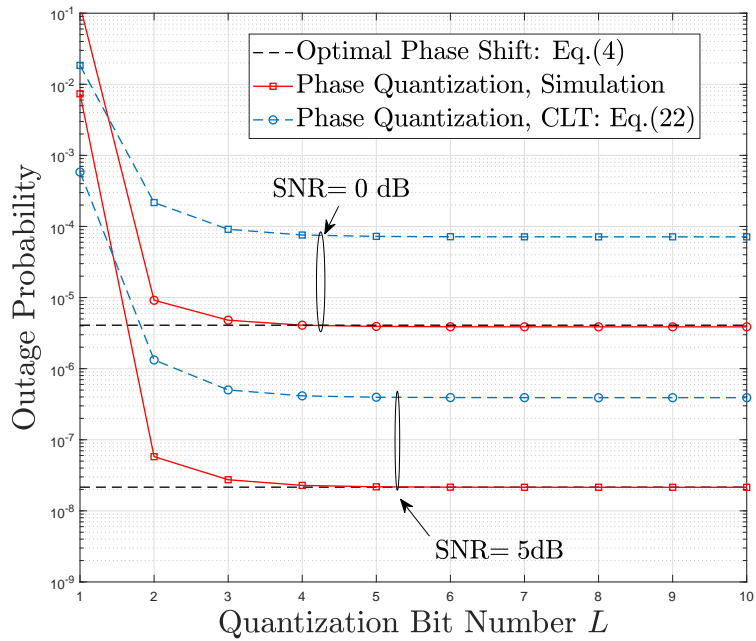}
\caption{Outage probability vs. $L$ over Nakagami-$m$ fading ($m=1.5$) for different values of the average SNR and $N=4$.}
\end{figure}

Figure~4 shows the outage probability  as a function of the number of quantization bits for the phase shifts of the RIS. Figure~4 further corroborates the performance trends illustrated in Fig.~3. In particular, we observe that $L=3$ provides an outage probability that is close to the setup in the absence of phase noise. Furthermore, Fig.~4 confirms that the CLT does not provide reliable estimates of the outage probability in the considered setup.

\section{Conclusion}
In this paper, we  have introduced a comprehensive analytical framework for analyzing the outage probability and diversity order of RIS-assisted  communication systems over generalized fading channels and in the presence of phase noise. The proposed approach leverages the analytical formalism of the Fox's H functions. We have substantiated, over generalized fading channels, that the central limit theorem is not, in general, a suitable approach for analyzing the diversity order of RIS-aided systems. Therefore, we have introduced a new analytical approach for computing the diversity order in the absence of phase noise, and we have identified sufficient conditions for ensuring that the full diversity order is achieved in the presence of phase noise. In particular, we have proved that RIS-assisted communications achieve the full diversity order provided that the absolute difference between pairs of phase errors is less than $\pi/2$. The obtained findings are shown to be in agreement and to generalize previous results available in the literature.

\section{Appendix A}

For sufficiently large $N$, the distribution of $H_b=\frac{1}{N}H$ tends to a non-circularly
symmetric complex Gaussian RV where $X={\rm{Re}}\{H\}=\frac{1}{N}\sum_{i=1}^{N}|h_i||g_i|\cos(\theta_i)$ and $Y={\rm {Im}}\{H\}=\frac{1}{N}\sum_{i=1}^{N}|h_i||g_i|\sin(\theta_i)$ are approximately Gaussian due to the CLT. In particular, $X\sim {\cal N}(\nu, \sigma_X^{2})$ and $Y\sim {\cal N}(0,  \sigma_Y^{2})$, where, by using the second-order statistic computation method in \cite{stat}, we obtain $\nu={\cal K}_1 \Lambda_1\Lambda_2$,  $\sigma^{2}_X=\frac{1}{2N}\left(1+ {\cal K}_2-2{\cal K}_1^{2}\Lambda_1^{2}\Lambda_2^{2}\right)$, and $\sigma_Y^{2}=\frac{1}{2N}\left(1-{\cal K}_2\right)$, where \begin{equation}
\Lambda_1=\mathbb{E}\{|h_i|\} ~ \text{and}~ \Lambda_2=\mathbb{E}\{|g_i|\},
\label{mean}
\end{equation}
which, using (\ref{h}), is computed using the Mellin transform of the Fox's H function \cite[Eq. (2.8)]{mathai}\footnote{For instance, we obtain $\Lambda_i=\frac{\sqrt{\pi}}{2}, i=1,2$, for Rayleigh fading,  $\Lambda_i=\frac{\Gamma(m_i+\frac{1}{2})}{\sqrt{m_i}\Gamma(m_i)}$ for Nakagami-$m$ fading, and $\Lambda_i=\sqrt{\frac{4}{\pi (K_i+1)}}{}_1F_1\left(-\frac{1}{2},1, -K_i\right)$ in Rician fading, respectively.}.

Moreover, since the considered
phase errors distributions are symmetric around their mean value that is equal to zero, we obtain $\mathbb{E}\{X Y\}=0$. Hence, $X$ and $Y$ are uncorrelated RVs, and, hence, as normal RVs, independent.
Accordingly, $|H_b|^{2}$ is the sum of a
scaled non-central chi-squared RV $X^{2}$ and a gamma variable $Y^{2}$, which are mutually independent. Thus, we have
 \begin{eqnarray}
\!\!{\cal L}f_{|H_b|^{2}}(s)={\cal L}f_{X^{2}}(s){\cal L}f_{Y^{2}}(s) =\frac{e^{-\frac{\nu^{2} s}{1+2\sigma_X^{2} s}}}{\sqrt{1+2\sigma_X^{2}  s}\sqrt{1+2\sigma_Y^{2}s}}.
\label{eqs}
\end{eqnarray}

By applying the same steps as in (\ref{LMGF}), i.e., by computing the inverse Laplace transform of (\ref{eqs}), the distribution of $|H_b|^{2}$ can be formulated as
\begin{eqnarray}
{\rm P}\left(|H_b|^{2}<t\right)&=&{\cal L}^{-1}\left\{s^{-1}{\cal L}f_{|H_b|^{2}}(s), t \right\}\nonumber \\&\overset{(a)}{=}&
\sum_{k=0}^{\infty} \frac{(-1)^{k}\left(\nu^{2} \right)^{k}}{k!}{\cal L}^{-1}\left\{\frac{s^{k-1}}{\left(1+2\sigma_X^{2}  s\right)^{k+\frac{1}{2}}\left(1+2\sigma_Y^{2} s\right)^{\frac{1}{2}}}, t \right\},
\label{l1}
\end{eqnarray}
where $(a)$ follows by using $e^{-x}=\sum_{k=0}^{\infty} \frac{(-1)^{k}x^{k}}{k!}$.

Substituting $(1+x)^{-a}=\frac{1}{\Gamma(a)}\int_{{\cal L}}\Gamma(s)\Gamma(a-s) x^{-s} ds$ in (\ref{l1}), and applying and applying ${\cal L}^{-1}\{s^{-a};x\}=\frac{x^{a-1}}{\Gamma(a)}$ \cite[Eq. (2.15)]{mathai}, the distribution of  $|H_b|^{2}$ can be formulated as
\begin{eqnarray}
\!\!\!\!{\rm P}\left(|H_b|^{2}<t\right)&=&
\frac{1}{\pi (2 \pi w)^{2}} \int_{{\cal C}_1} \int_{{\cal C}_2} \frac{\Gamma(u_1)\Gamma(u_2)}{{(2\sigma_X^{2})}^{u_1}{(2\sigma_Y^{2})}^{u_2}} \frac{\Gamma\left(\frac{1}{2}\!-\!u_1\right) \Gamma\left(\frac{1}{2}\!-\!u_2\right)}{\Gamma(1+u_1+u_2)}\nonumber \\ &&\!\!\!\!\!\!\!\!\!\!\!\!\!\!\!\!\!\!\!\!\!\!\!\!\!\!\!\!\!\!\!\!\! \hspace{3cm} \times  t^{u_1\!+\!u_2\!}{\rm {}_{2}F_{1}}\left(\frac{1}{2}-u_1, -u_1-u_2,\frac{1}{2}, \frac{\nu^{2}}{t}\right)du_1 du_2,
\label{per}
\end{eqnarray}
where  ${\rm {}_{2}F_{1}}(\cdot)$  denotes  the Gauss hypergeometric function \cite{grad}. Using the representation of the ${\rm {}_{2}F_{1}}(\cdot)$ hypergeometric function   in terms of Mellin-Barnes integrals \cite{mathai}, we obtain (36) shown at the top of the next page.
 \begin{figure*}
\begin{eqnarray}
\!\!\!\!{\rm P}\left(|H_b|^{2}<t\right)&=&
\frac{1}{\sqrt{\pi} (2 \pi w)^{3}} \\ && \!\!\!\!\!\!\!\!\!\!\!\!\!\!\!\!\!\!\!\!\!\!\!\!\!\!\!\!\!\!\! \hspace{-1.5cm} \times \int_{{\cal C}_1} \int_{{\cal C}_2}  \int_{{\cal C}_3} \frac{\Gamma(u_1)\Gamma(u_2)\Gamma(u_3)}{{(\frac{2\sigma_X^{2}}{\nu^2 t})}^{u_1}{(\frac{2\sigma_Y^{2}}{
\nu^2 t})}^{u_2}(\nu^{2})^{ u_1+u_2+u_3}} \frac{\Gamma\left(\frac{1}{2}\!-\!u_1-u_3\right)\Gamma(-u_1-u_2-u_3) \Gamma\left(\frac{1}{2}\!-\!u_2\right)}{\Gamma(1+u_1+u_2)\Gamma\left(\frac{1}{2}-u_3\right)\Gamma\left(-u_1-u_2\right)} (-t)^{u_3\!}du_1 du_2 du_3 \nonumber
\label{perint}
\end{eqnarray}
\hrulefill
\end{figure*}\\
From (\ref{snoise}), the outage probability for large $N$ is obtained as \begin{equation}
   \Pi(\rho,N) ={\rm P}\left(|H_b|^{2}<\frac{\gamma_{th}}{N^{2} \rho}\right).
   \end{equation}
However, in the obtained current form, the distribution of $|H_b|^{2}$ in (\ref{perint}) involves an undermined form  in the high-SNR regime. When $t\rightarrow 0$, more precisely, we have
\begin{equation}
\lim_{N \rightarrow \infty}\lim_{ t\rightarrow 0}\frac{\nu^{2} t}{\sigma_X^{2}} =\lim_{N \rightarrow \infty}\lim_{ t\rightarrow 0}\frac{\nu^{2}t}{\sigma_Y^{2}}=0 \times \infty.
\end{equation}

To circumvent this, we use the Euler-Gauss limit \cite{grad} for $Z\in \{\frac{\nu^{2}}{2\sigma^{2}_X}, \frac{\nu^{2}}{2\sigma^{2}_Y} \}$ as
\begin{equation}
Z^{s}\underset{N\gg1}{\simeq} \frac{\Gamma(Z-s)}{\Gamma(Z-2s)}.
\label{apn}
\end{equation}
Based on \eqref{apn}, we obtain (\ref{perint1}),  shown at the top of this page.
 \begin{figure*}
\begin{eqnarray}
\!\!\!\!{\rm P}\left(|H_b|^{2}<t\right)&=&
\frac{1}{\sqrt{\pi} (2 \pi w)^{3}}\nonumber \\ && \!\!\!\!\!\!\!\!\!\!\!\!\!\!\!\!\!\!\!\!\!\!\!\!\times \int_{{\cal C}_1} \int_{{\cal C}_2}  \int_{{\cal C}_3} \Gamma(u_1)\Gamma(u_2)\Gamma(u_3) \frac{\Gamma\left(\frac{1}{2}\!-\!u_1-u_3\right)\Gamma(-u_1-u_2-u_3) \Gamma\left(\frac{1}{2}\!-\!u_2\right)}{\Gamma(1+u_1+u_2)\Gamma\left(\frac{1}{2}-u_3\right)\Gamma\left(-u_1-u_2\right)} \nonumber \\ &&
\frac{\Gamma\left(\frac{\nu^{2}}{2 \sigma^{2}_X}-u_1\right)}{\Gamma\left(\frac{\nu^{2}}{2 \sigma^{2}_X}-2u_1\right)}\frac{\Gamma\left(\frac{\nu^{2}}{2 \sigma^{2}_Y}-u_2\right)}{\Gamma\left(\frac{\nu^{2}}{2 \sigma^{2}_Y}-2u_2\right)}
\left(\frac{t}{\nu^{2}}\right)^{u_1+u_2}\left(\frac{-t}{\nu^{2}}\right)^{u_3}du_1 du_2 du_3.
\label{perint1}
\end{eqnarray}
\hrulefill
\end{figure*}
By applying \cite[eq. (A.1)]{mathai} to (\ref{perint1}) and using the identity $\Gamma\left(\frac{1}{2}-u_3\right)=\frac{\Gamma(-2 u_3) 2^{2u_3+1}\sqrt{\pi}}{\Gamma(-u_3)}$, we obtain (\ref{outsig}) after some manipulations.

If the phase noise has a uniform distribution over $[-\pi, \pi]$, we have ${\rm {}_{2}F_{1}}\left(\frac{1}{2}-u_2, -u_1-u_2, 0 \right)=1$ in (\ref{per}), since $\nu=0$. Substituting $\sigma^{2}_X=\sigma^{2}_Y=\frac{1}{2N}$, and using the Euler-Gauss limit \cite{grad}
\begin{equation}
{N}^{s}\underset{N\gg1}{\simeq} \frac{\Gamma(N-s)}{\Gamma(N-2s)},
\label{apn2}
\end{equation}
we obtain
\begin{eqnarray}
\!\!\!\!{\rm P}\left(|H_b|^{2}<t\right)&=&
\frac{1}{\pi (2 \pi w)^{2}}\nonumber \\ && \!\!\!\!\!\!\!\!\!\!\!\!\!\!\!\!\!\!\!\!\!\!\!\! \hspace{-1cm}\times \int_{{\cal C}_1} \int_{{\cal C}_2} \Gamma(u_1)\Gamma(u_2) \frac{\Gamma\left(\frac{1}{2}\!-\!u_1\right) \Gamma\left(\frac{1}{2}\!-\!u_2\right)}{\Gamma(1+u_1+u_2)} \frac{\Gamma(N-u_1)}{\Gamma(N-2u_1)}\frac{\Gamma(N-u_2)}{\Gamma(N-2u_2)} t^{u_1\!+\!u_2\!}du_1 du_2,
\label{per22}
\end{eqnarray}
which leads to (\ref{outun}) with the aid of \cite[eq. (A.1)]{mathai}. This completes the proof.

\end{document}